\documentclass[11pt]{article}
\usepackage{jheppub}
\usepackage{amsmath,bbm}

\makeatletter
\def\@fpheader{\relax}
\makeatother

\usepackage{graphicx}
\usepackage{caption}
\usepackage{subcaption}
\usepackage{verbatim}

\usepackage{tocvsec2}
\settocdepth{section}
\toccontinuoustrue

\title{\huge Entropy of a subalgebra of observables and the geometric entanglement entropy}

\author[a\;]{Eugenio Bianchi,}
\emailAdd{ebianchi@gravity.psu.edu}
\affiliation[a]{Institute for Gravitation and the Cosmos \& Physics Department,\\ Penn State, University Park, PA 16802, USA}

\author[b\;]{Alejandro Satz\,}
\emailAdd{asatz@sarahlawrence.edu}
\affiliation[b]{Sarah Lawrence College, Bronxville, NY 10708, USA}

\abstract{
The geometric entanglement entropy of a quantum field in the vacuum state is known to be divergent and, when regularized, to scale as the area of the boundary of the region. Here we introduce an operational definition of the entropy of the vacuum restricted to a region: we consider a subalgebra of observables that has support in the region and a finite resolution. We then define the entropy of a state restricted to this subalgebra. For Gaussian states, such as the vacuum of a free scalar field, we discuss how this entropy can be computed. In particular we show that for a spherical region we recover an area law under a suitable refinement of the subalgebra.
}

\begin{document}
\maketitle
\newpage

\section{Introduction}
In quantum field theory, the geometric entanglement entropy is a quantity associated to a pure state of the field---typically the vacuum---and a region of space \cite{Sorkin:2014kta,Bombelli, Srednicki}. This quantity has proven to be a fundamental tool for investigating properties of quantum fields in various settings, ranging from the study of quantum fields in the presence of black hole horizons \cite{Solodukhin:2011gn}, characterizing ground states of many-body systems \cite{Amico:2007ag}, identifying new phases of quantum matter \cite{zeng2019quantum}, proving conjectures on the running of coupling constants \cite{Casini:2015woa}, and exploring the quantum nature of space-time geometry \cite{Jacobson:1995ab, VanRaamsdonk:2010pw, Bianchi:2012ev, Ryu:2006bv, Jacobson:2015hqa}.

In order to define the geometric entanglement entropy in quantum field theory, an ultraviolet cut-off is needed. The origin of this divergence is the short-distance correlations at space-like separation present in all regular states of a quantum field \cite{Kay:1988mu}. A standard procedure involves a discretization \cite{CasiniHuertaRev}: the field is put on a lattice and the state is defined to be, for instance, the ground state of the lattice Hamiltonian. The entanglement entropy is then computed before the continuum limit is taken. This is the method that was originally used to show that the geometric entanglement entropy of the Minkowski vacuum state satisfies an area law \cite{Sorkin:2014kta,Bombelli, Srednicki, CasiniHuertaRev}. The result is reproduced also by using other regularization methods such as the brick-wall cutoff \cite{tHooft:1984kcu}, Pauli-Villars regulators \cite{Demers:1995dq}, conical-defect methods based on the replica trick \cite{Susskind:1994sm, Callan:1994py}, holographic methods where the cut-off is encoded in the distance from the AdS boundary \cite{Ryu:2006bv}, and the use of the mutual information to introduce a ``safety-corridor'' between the region and its complement \cite{Casini:2008wt}.

From an operational point of view, entropy is a measure of the uncertainty of outcomes of measurements \cite{NielsenChuang}. Adopting this perspective for the geometric entanglement entropy can be fruitful, especially in view of prospects of direct measurements of the entanglement entropy in condensed matter systems \cite{islam}.
The algebraic approach to quantum field theory \cite{Haag, Halvorson:2006wj, Hollands:2017dov} provides an efficient language to formalize this notion. In this setting, a subsystem $R$ is identified by a restricted set of measurements, i.e. a subalgebra $\mathcal{A}_R\subset \mathcal{A}$ of the algebra of observables of the system. The entanglement entropy $S_R(|\psi\rangle)$ is the entropy of the state $|\psi\rangle$ restricted to the subalgebra of observables $\mathcal{A}_R$ \cite{OhyaPetz}. For systems with a finite number of degrees of freedom and a subalgebra that selects some of its degrees of freedom, this definition coincides with the standard procedure which involves the computation of a reduced density matrix and the computation of its von Neumann entropy \cite{NielsenChuang}. On the other hand, in a field theoretic setting with infinitely many degrees of freedom, the algebraic setting provides a useful generalization.

In the algebraic setting, the divergent value of the geometric entanglement entropy is rooted in the properties of observables localized in a region of space. For a free scalar field in a canonical setting for instance, we can consider the algebra of observables $\mathcal{A}_R$ generated by the field $\varphi(\vec{x})$ and its conjugate variable $\pi(\vec{x})$, with $\vec{x}\in R$. Clearly, observables in the region $R$ and observables in its complement $\bar{R}$ commute, i.e. $[\mathcal{A}_R,\mathcal{A}_{\bar{R}}]=\{0\}$. However, this fact is not sufficient to guaranty statistical independence of the two subalgebras, i.e. $\mathcal{A}\neq \mathcal{A}_R\times\mathcal{A}_{\bar{R}}$. Technically, one says that $\mathcal{A}_R$ is of type III \cite{Haag, Halvorson:2006wj, Hollands:2017dov}. A consequence of the lack of statistical independence is that there are no pure states on $\mathcal{A}_R$ and no absolute notion of how to set the zero of the geometric entanglement entropy. The standard procedures used to make sense of the geometric entanglement entropy either modify the theory (for instance via a discretization), or focus on quantities that do not directly measure the entropy of observables in a region (such as the mutual information with a safety corridor).

In this paper we adopt an operational approach where one identifies what an experiment can measure in principle. To this effect, we consider a finite-dimensional subalgebra of observables, defined by smearing the field operator (and its conjugate momentum) with a finite set of smearing functions. The resulting finite set of observables are meant to represent observables one might have experimental access to, such as the average value of the field (or some component of it in a mode expansion) in a finite spatial region. In Section \ref{sec:Gaussian} we provide the general  definition of such a subalgebra. Moreover we show how, in the case where the field is in a Gaussian state, we can explicitly define the Von Neumann entropy of the subalgebra. This entropy measures the entanglement between the selected observables and the other modes of the field. Unlike the geometric entropy, this quantity is well-defined and finite by construction. In Section \ref{sec:vacuum} we provide examples in which the field is smeared with Gaussian functions in a region of size $R$, providing explicit computations of the entanglement entropy associated to these observables. In Section \ref{sec:sphere} we introduce and define an observable subalgebra adapted to a spherical region that simplifies the definition and evaluation of the entanglement entropy in the limit to the full 
Type III algebra $\mathcal{A}_R$. In Section \ref{sec:observables} we explicitly compute the entropy in this limit, showing that---as expected---it is divergent, and that the leading divergent term reproduces the familiar area law for the geometric entropy. This confirms that our definition captures a finite version of the geometric entropy which, unlike its standard regularizations, is associated with a concrete set of field observables and not with an artificially cutoff of the dynamics of the theory. Section \ref{sec:discussion} contains a summary and discussion of the main results.

\section{Gaussian states, subalgebras of observables and entanglement}\label{sec:Gaussian}
We consider a free scalar field in Minkowski space. In the canonical formulation one starts with a fixed-time slice, with the field operator $\phi(\vec{x})$ and the momentum operator $\pi(\vec{x})$ satisfying the equal-time canonical commutation relations:
\begin{equation}
[\phi(\vec{x}),\phi(\vec{y})]\;=\;0\,,\qquad [\pi(\vec{x}),\pi(\vec{y})]\;=\;0\,,\qquad [\phi(\vec{x}),\pi(\vec{y})]\;=\;\mathrm{i}\,\delta(\vec{x}-\vec{y})\,.
\end{equation}
It is useful to pack the canonical couple into a single field with two components
\begin{equation}
 \chi^r(\vec{x})=
 \left(
\begin{array}{c}
\phi(\vec{x})\\[2pt]
\pi(\vec{x})
\end{array}
\right)
\,,\qquad r=1,2\,.
\end{equation}
The commutation relations take then the form
\begin{equation}
[\,\chi^r(\vec{x}),\chi^s(\vec{y})\,]\;=\;\mathrm{i}\,\sigma^{rs}\;\delta(\vec{x}-\vec{y})\qquad\mathrm{with}\qquad
\sigma^{rs}=\left(
\begin{array}{rr}
\;0 &\;\;1\\[2pt]
-1&0
\end{array}
\right).
\label{eq:commutation-chi}
\end{equation}
The algebra $\mathcal{A}$ of observables of the system consists of linear combinations of symmetrized products of smeared fields
\begin{equation}
\chi_f=\int f_r(\vec{x})\;\chi^r(\vec{x})\;d\vec{x}
\label{eq:linear-obs}
\end{equation}
where $f_r(\vec{x})$ is a smooth function. The Hilbert space of the system is the Fock space $\mathcal{H}$ built over the Minkowski vacuum $|0\rangle$.\footnote{The annihilation operator $a(\vec{k})$ which defines the Minkowski vacuum, $a(\vec{k})|0\rangle=0$ for all $\vec{k}$,  is a linear combination of the form (\ref{eq:linear-obs}), i.e.,  $a(\vec{k})=\omega(\vec{k})\,\phi(\vec{k})+\mathrm{i}\,\pi(\vec{k})$
where $\omega(\vec{k})=\sqrt{\vec{k}^2+m^2}$, $m$ is the mass of the field and $\phi(\vec{k})$, $\pi(\vec{k})$ are the Fourier transforms of the field and momentum operators.} Given a state $|\psi\rangle\in \mathcal{H}$ we can compute the equal-time $n$-point correlation functions. In particular, a Gaussian state has correlation functions
\begin{align}
\langle\psi|\, \chi^r(\vec{x}) \,|\psi\rangle\;=&\;\;0 \label{eq:1-corr} \\[.5em]
\langle\psi|\, \chi^r(\vec{x}) \,\chi^s(\vec{y}) \,|\psi\rangle\;=&\;\;\frac{C^{rs}(\vec{x},\vec{y})\;+\;\mathrm{i}\,\sigma^{rs}\;\delta(\vec{x}-\vec{y})}{2}\;,\label{eq:2-corr}
\end{align}
and all higher-$n$ correlation functions determined by their Wick relations in terms of the $2$-point correlation function. The antisymmetric part of the correlation function is fixed by the commutation relations (\ref{eq:commutation-chi}). The symmetric part is given by
\begin{equation}
C^{rs}(\vec{x},\vec{y})=2\left(
\begin{array}{cc}
\langle\psi|\phi(\vec{x}) \,\phi(\vec{y}) |\psi\rangle& \langle\psi|\frac{\phi(\vec{x})\pi(\vec{y})+\pi(\vec{y})\,\phi(\vec{x})}{2} |\psi\rangle\\[1em]
 \langle\psi|\frac{\phi(\vec{x})\pi(\vec{y})+\pi(\vec{y})\,\phi(\vec{x})}{2} |\psi\rangle\;\;&\langle\psi|\pi(\vec{x}) \,\pi(\vec{y}) |\psi\rangle
\end{array}
\right).
\end{equation}
For a Gaussian state, the expectation value of any observable can be expressed in terms of the symmetric correlation function $C^{rs}(\vec{x},\vec{y})$ and the canonical commutator (\ref{eq:commutation-chi}).\\

In concrete situations, an experiment has access only to a subset of all the possible measurements that can be performed on the state $|\psi\rangle$. This subset of measurements is described by a subalgebra of observables and defines a subsystem. We consider the subsystem $A$ identified by the subalgebra  $\mathcal{A}_A\subset \mathcal{A}$ generated by $2N_A$ linear observables
\begin{equation}
\xi^a=\int f^a_r(\vec{x}\,)\,\chi^r(\vec{x}\,)\;d^{3}\vec{x}\;,\qquad a=1,\ldots,2N_A
\end{equation}
where $f^a_r(\vec{x})$ are a set of smearing functions satisfying the following constraint: we require that 
\begin{equation}
[\xi^a,\xi^b]=\mathrm{i}\,\Omega^{ab} \qquad\textrm{with}\;\, \Omega^{ab}\; \textrm{a symplectic structure on}\; \mathbb{R}^{2N_A}.
\label{eq:symplectic}
\end{equation}
This requirement results in the condition that the smearing functions $f^a_r(\vec{x})$ define a real $2N_A\times 2N_A$ antisymmetric  matrix 
\begin{equation}
\Omega^{ab}=\int f^a_r(\vec{x}\,)f^b_s(\vec{x}\,)\,\sigma^{rs}\,d\vec{x}
\end{equation}
which is invertible. When this condition is satisfied, the couple $(\mathbb{R}^{2N_A},\Omega^{ab})$ is a symplectic vector space and the algebra generated by the $2N_A$ linear observables $\xi^a$ is the Weyl algebra $\mathcal{A}_A=\mathrm{Weyl}(2N_A,\mathbb{C})$. As a result the subsystem $A$ is an ordinary quantum mechanical system with the Hilbert space $\mathcal{H}_A$ of a finite number $N_A$ of bosonic degrees of freedom.

The $n$-point correlation functions for the subsystem can be computed directly from Eq.~(\ref{eq:1-corr}) and (\ref{eq:2-corr}). In particular the expectation value of the linear observable $\xi^a$ vanishes, $\langle\psi|\, \xi^a \,|\psi\rangle=0$, and the correlations functions of the subsystem are
\begin{equation}
\langle\psi|\, \xi^a\,\xi^b \,|\psi\rangle\;=\;\frac{G^{ab}+\mathrm{i}\;\Omega^{ab}}{2}
\label{eq:corr-psi}
\end{equation}
where
\begin{equation}
G^{ab}=\int f^a_r(\vec{x}\,)f^b_s(\vec{y}\,)\,C^{rs}(\vec{x},\vec{y}\,)\,d\vec{x}\,d\vec{y}\,.
\end{equation}
is a real $2N_A\times 2N_A$ symmetric  matrix. From the definition (\ref{eq:corr-psi}) it is immediate to prove that the matrices $G^{ab}$ and $\Omega^{ab}$ have the follow properties:
\begin{align}
&G^t=G\,,\quad\; \Omega^t=-\Omega\,,\label{eq:sym}\\
&G>0\,,\qquad \exists \;\Omega^{-1}\,,\\
&G+\mathrm{i}\,\Omega\geq 0\,,
\label{eq:G+O}
\end{align}
where we have adopted a matrix index-free notation for $G^{ab}$ and $\Omega^{ab}$ and defined  $G^t$ as the matrix transpose of $G$. The existence of the inverse $\Omega^{-1}$ follows from the condition (\ref{eq:symplectic}). To prove that $G>0$ we can consider the expectation value of the positive Hermitian operator $\mathcal{O}=v_a v_b\, \xi^a\xi^b$ with $v_a\in \mathbb{R}^{2N_A}$. We have $0\leq \langle\psi|\mathcal{O}|\psi\rangle=\frac{1}{2}G^{ab}v_av_b$ for all $v_a$ which implies that $G^{ab}$ is positive-definite. Similarly for the condition (\ref{eq:G+O}) we can consider 
the positive Hermitian operator $\mathcal{O}=z^*_a z^{\phantom{*}}_b \xi^a\xi^b$ with $z_a\in \mathbb{R}^{2N_A}$. We have $0\leq \langle\psi|\mathcal{O}|\psi\rangle=\frac{1}{2}(G^{ab}+\mathrm{i}\,\Omega^{ab})z^*_a z^{\phantom{*}}_b$ for all $z_a$ which implies that the Hermitian matrix $G+\mathrm{i}\,\Omega$ has non-negative eigenvalues. It is also useful to define the $2N_A\times 2N_A$ real matrix
\begin{equation}
J_{\!A}=G\,\Omega^{-1}\,.
\label{eq:Jdef}
\end{equation}
As a consequence of Eq.~(\ref{eq:G+O}), the matrix $\mathrm{i} J_{\!A}$ has real eigenvalues which appear in pairs of opposite sign and magnitude equal or larger than one,
\begin{equation}
\mathrm{Eig}(\,\mathrm{i} J_{\!A})\,=\,\pm\, \nu_i\,,\qquad \textrm{with}\quad \nu_i\geq 1\quad \textrm{and}\quad i=1,\ldots, N_A\,.
\label{eq:EigJ}
\end{equation}
The matrix $J_{\!A}$ is a restricted complex structure: it is the complex structure of the Gaussian state $|\psi\rangle$ restricted to the subalgebra $\mathcal{A}_A$. Here we use the linear symplectic methods developed for Gaussian states in \cite{holevo2013quantum,weedbrook2012gaussian,adesso2014continuous,Bianchi:2017kgb,Bianchi2017kahler}.

For a Gaussian state $|\psi\rangle$ of the quantum field, the matrices $G^{ab}$ and $\Omega^{ab}$ describe completely all the properties of the subsystem identified by the observables in the subalgebra $\mathcal{A}_A$ generated by the linear operators $\xi^a$. We can
in fact introduce a mixed density matrix $\rho$ defined on the Hilbert space $\mathcal{H}_A$ of a bosonic system with $N_A$ degrees of freedom,
\begin{equation}
\rho=\frac{e^{-q_{ab}\xi^a\xi^b}}{Z}\qquad\textrm{with}\qquad q_{ab}=\big(\mathrm{i}\,\Omega^{-1}\;\mathrm{arcoth}(\mathrm{i}J_{\!A})\big){}_{ab}
\label{eq:rho}
\end{equation}
where $Z$ is such that $\mathrm{Tr}(\rho)=1$
. The real symmetric $2N_A\times 2N_A$  matrix  $q_{ab}$ is positive-definite as a consequence of Eq.~(\ref{eq:EigJ}).
For all observables $\mathcal{O}_A\in \mathcal{A}_A\subset \mathcal{A}$ we have
\begin{equation}
\langle\psi|\mathcal{O}_A|\psi\rangle=\mathrm{Tr}(\mathcal{O}_A\,\rho)\;.
\end{equation}
In particular $\mathrm{Tr}(\xi^a\rho)=0$ and $\mathrm{Tr}(\xi^a\,\xi^b\,\rho)=\frac{1}{2}(G^{ab}+\mathrm{i}\;\Omega^{ab})$, therefore reproducing the correlation function (\ref{eq:corr-psi}). Defining the orthonormal basis $|n_1,\ldots, n_{N_A}\rangle$ of $\mathcal{H}_A$ which diagonalizes the quadratic operator $q_{ab}\,\xi^a\xi^b$ appearing in the exponent of Eq.~(\ref{eq:rho}), we find that the density matrix can be expressed as
\begin{equation}
\rho=\sum_{n_i=0}^\infty\left(\;\prod_{i=1}^{N_A} \frac{2}{\nu_i+1}\Bigg(\frac{\nu_i -1}{\nu_i +1}\Bigg)^{\!\!n_i}\;\right)\;|n_1,\ldots, n_{N_A}\rangle\langle n_1,\ldots, n_{N_A}|
\label{eq:rhonn}
\end{equation}
where $\nu_i$ are the positive eigenvalues of $\mathrm{i}J_{\!A}$ defined in Eq.~(\ref{eq:EigJ}).

The density matrix $\rho$ provides a representation of the restriction of the Gaussian state $|\psi\rangle$ to the subalgebra of observables $\mathcal{A}_A$. While the state $|\psi\rangle$ is pure and therefore has zero entropy, its restriction to the subalgebra $\mathcal{A}_A$ results in an von Neumann entropy
\begin{equation}
S_A(|\psi\rangle)= -\mathrm{Tr}(\rho\log\rho)=\sum_{i=1}^{N_A} s(\nu_i)
\label{eq:sentropy}
\end{equation}
where $s(\nu)$ is the function
\begin{equation}
s(\nu)= \;\frac{\nu+1}{2}\log\frac{\nu+1}{2}\;-\frac{\nu-1}{2}\log\frac{\nu-1}{2}\,.
\label{eq:snu}
\end{equation}
The origin of the entropy $S_A(|\psi\rangle)$ is the entanglement between the restriction of the state $|\psi\rangle$ to the subalgebra $\mathcal{A}_A$ and its complement.

The algebra of observables describing the rest of the system is given by the set of all operators that commute with all operators in $\mathcal{A}_A$, also known as the commutant $\mathcal{A}'_A$,
\begin{equation}
\mathcal{A}'_A\equiv \{\mathcal{O}\in \mathcal{A}\;|\;[\mathcal{O}_A,\mathcal{O}]=0\;\;\textrm{for all}\;\; \mathcal{O}_A\in \mathcal{A}_A\}.
\end{equation}
In our case, the subalgebra  $\mathcal{A}_A$ has a trivial center,  i.e. $\mathcal{A}_A\cap\mathcal{A}'_A=\mathbbm{1}$. As a result $\mathcal{A}_A$ is a factor, the complement of the subsystem $A$ is the subsystem $B$ defined by the subalgebra $\mathcal{A}_B=\mathcal{A}'_A$ and we have the decomposition $\mathcal{A}=\mathcal{A}_A\otimes \mathcal{A}_B$. Moreover, as the subalgebra  $\mathcal{A}_A=\mathrm{Weyl}(2N_A,\mathbb{C})$ is finitely generated, it is of type $\mathrm{I}$ and the Hilbert space of the system decomposes in the tensor product $\mathcal{H}=\mathcal{H}_A\otimes \mathcal{H}_B$ \cite{Haag}. Therefore the entropy of the subalgebra $\mathcal{A}_A$ is the entanglement entropy between the subsystems with Hilbert space $\mathcal{H}_A$ and $\mathcal{H}_B$.

\section{Vacuum entropy of observables with Gaussian smearing}
\label{sec:vacuum}
Let us consider the vacuum state $|0\rangle$ of a free scalar field in $4d$ Minkowski space. The symmetric part of the equal-time correlation function is given by
\begin{equation}
C^{rs}(\vec{x},\vec{y})=2\left(
\begin{array}{cc}
\langle 0 |\phi(\vec{x}) \,\phi(\vec{y}) |0\rangle& 0\\[.5em]
0 &\langle 0|\pi(\vec{x}) \,\pi(\vec{y}) |0\rangle
\end{array}
\right)=\int\!\frac{d^3\vec{k}}{(2\pi)^3}\left(
\begin{array}{cc}
\omega(\vec{k})^{-1}&0\\[.5em]
0&\omega(\vec{k})
\end{array}
\right) \;e^{-\mathrm{i}\vec{k}\cdot(\vec{x}-\vec{y})}\,,
\end{equation}
where $\omega(\vec{k})=\sqrt{\vec{k}^2+m^2}$ and $m$ is the mass of the field. We consider a subalgebra of observables generated by a Gaussian smearing of the field $\phi(\vec{x})$ and the momentum $\pi(\vec{x})$ over a region of size $R$,
\begin{align}
\Phi_R=&\,\frac{1}{(2\pi)^{3/2}R^3}\int d^3\,x\, e^{-r^2/2R^2}\,\phi(\vec{x})\;, \label{eq:PhiR}\\[1em]
\Pi_R=&\,\frac{1}{(2\pi)^{3/2}R^3}\int d^3\,x\, e^{-r^2/2R^2}\,\pi(\vec{x})\,.
\label{eq:PiR}
\end{align}
In the limit $R\to0$, the observables  $\Phi_R$ and $\Pi_R$ reproduce the distributional operators $\phi(\vec{x})$ and $\pi(\vec{x})$ evaluated at the $\vec{x}=\vec{0}$. For finite $R$ they can be interpreted as what a detector with a finite resolution $R$ measures. The commutator of  $\Phi_R$ and $\Pi_R$ is
\begin{equation}
[\Phi_R,\Pi_R]=
\mathrm{i}\,\frac{1}{8\pi^{3/2}\,R^3}\,.
\label{eq:vacuumome}
\end{equation}
Defining the dimensionless variable $\mu=m R$, the vacuum variance of the smeared observables is
\begin{align}\label{eq:vacuumphi2}
&\langle 0|\Phi_R\,\Phi_R|0\rangle=\frac{\mu^2\; e^{\mu^2/2}}{2}\Big(K_1(\mu^2/2)-K_0(\mu^2/2)\Big)\;\frac{1}{(2\pi)^2\,R^2}\quad\overset{\mu\to 0}{\longrightarrow}\quad \frac{1}{(2\pi)^2\,R^2}\\[1em]
\label{eq:vacuumPi2}
&\langle 0|\Pi_R\,\Pi_R|0\rangle=\frac{\mu^2\; e^{\mu^2/2}}{2}K_1(\mu^2/2)\;\frac{1}{(2\pi)^2\,R^4}\quad\overset{\mu\to 0}{\longrightarrow}\quad \frac{1}{(2\pi)^2\,R^4}\\[1em]
&\label{eq:vacuumcros}\langle 0|\Phi_R\,\Pi_R+\Pi_R\,\Phi_R|0\rangle=0\,.
\end{align}
In the language of the previous section, $f^a_r(\vec{x})=\frac{1}{(2\pi)^{3/2}R^3}\, e^{-r^2/2R^2}\,\delta^a_r$. The components of the correlation matrix $G^{ab}$ are given by equations (\ref{eq:vacuumphi2}-\ref{eq:vacuumcros}), while the nontrivial components of the symplectic form matrix $\Omega^{ab}$ are given by $\pm \frac{1}{i}[\Phi_R,\Pi_R]$ as given in (\ref{eq:vacuumome}). The restricted complex structure $J_{\!R}$ associated to our subalgebra is then computed by (\ref{eq:Jdef}) to be:
\begin{equation}
J_{\!R} = G\,\Omega^{-1} =\frac{1}{4\pi^{3/2} R^3}\left( \begin{array}{cc}
  0 &  -\langle 0|\Pi_R\,\Pi_R|0\rangle   \\
\langle 0|\Phi_R\,\Phi_R|0& 0\rangle \end{array} \right)\,.
\end{equation}

The positive eigenvalue of $\mathrm{i}J_{\!R}$ is $\nu(\mu)=2\big( \langle 0|\Phi_R\,\Phi_R| 0\rangle \langle 0|\Pi_R\,\Pi_R|0\rangle \big)^{1/2}$. The entropy $S(\mu)$ associated to this subalgebra is then given by (\ref{eq:snu}). 

\begin{figure}[t]
\begin{center}
\includegraphics[width=0.6\textwidth]{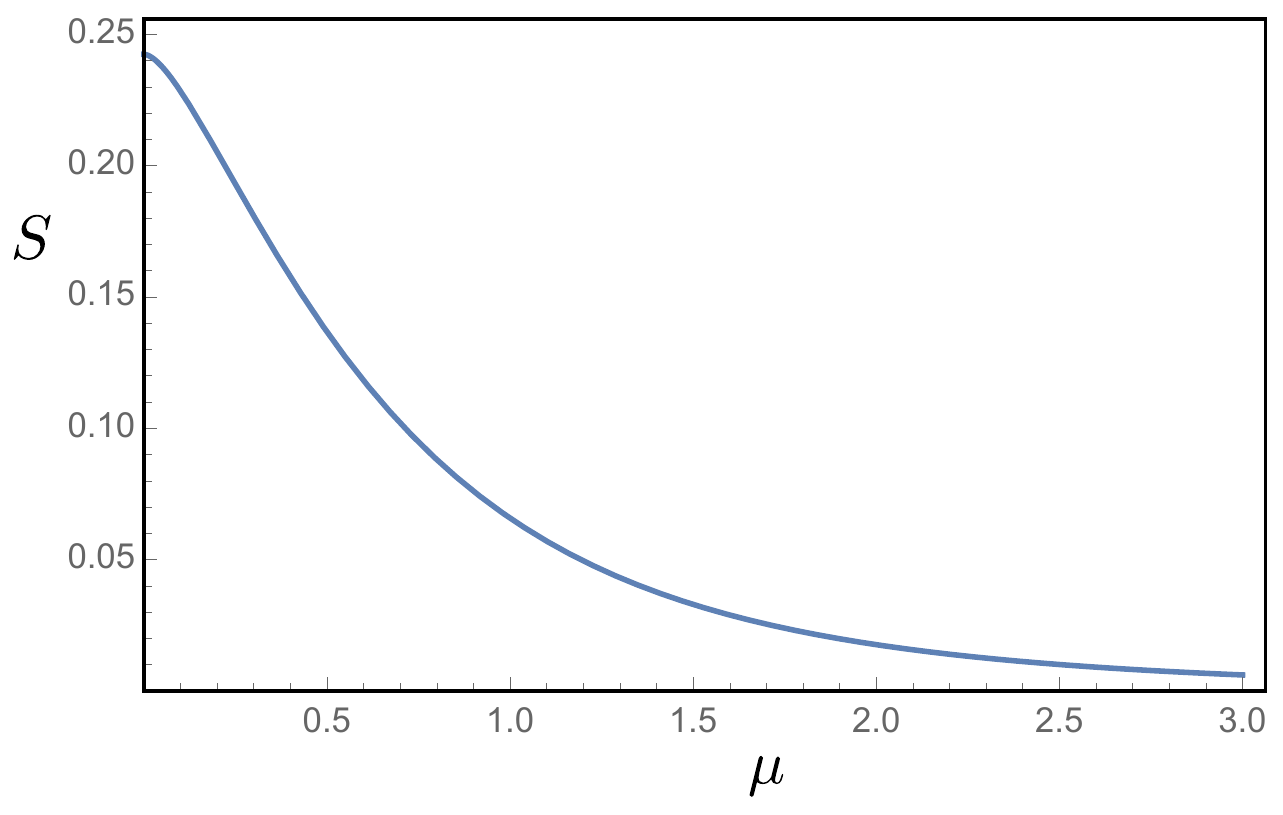}
\end{center}
\caption{Entropy of the $(\Phi_R,\Pi_R)$ subalgebra for a massive field in the Minkowski vacuum, as a function of $\mu=mR$ . }
\label{fig:massiveplot} 
\end{figure}

This entropy provides a measure of the entanglement between a single Gaussian-smeared degree of freedom (for the field and its momentum) on a region of size $R$, as defined by (\ref{eq:PhiR})--(\ref{eq:PiR}), and the degrees of freedom complementary to it. The entropy as a function of $\mu$ is plotted in Figure \ref{fig:massiveplot}. The large $\mu$ limit corresponds to the smeared measurements taking place over a region much larger than the Compton wavelength $m^{-1}$; hence, no information about fluctuations is registered, and the entropy vanishes. Accordingly, $\nu(\mu)\to 1$ at large $\mu$ indicating that the uncertainty relation is saturated. 

In the massless limit of the correlators, exhibited in (\ref{eq:vacuumphi2}) and (\ref{eq:vacuumPi2}), the positive eigenvalue of the restricted complex structure $\mathrm{i}J_{\!R}$ takes the value $\nu_0=2\pi^{-1/2}$. The entropy for the Gaussian-smeared observables of the massless scalar field takes the value $S_0 \approx 0.24$, seen as the $\mu\to 0$ limit in Figure \ref{fig:massiveplot}. This entropy is independent of the size $R$ of the region, reflecting the conformal invariance of the massless theory.

\subsection{Entropy of a larger subalgebra}

We consider now an extension of the previous subalgebra to include more information about the field's degrees of freedom in a region of size $R$. Focusing on the massless field for simplicity, we consider the subalgebra generated by $n$ pairs of smeared field and momentum observables $(\Phi^{k_1}_R,\Pi^{k_1}_R,\Phi^{k_2}_R,\Pi^{k_2}_R,\cdots \Phi^{k_n}_R,\Pi^{k_n}_R)$. The new set of observables is defined by:
\begin{equation}\label{eq:PhikR}
\Phi^k_R=\frac{\mathrm{e}^{k^2R^2/2}}{(2\pi)^{3/2}R^3}\int d^3x\, \mathrm{e}^{-r^2/2R^2}\,\frac{\sin(kr)}{kr}\,\phi(\vec{x})\,.
\end{equation}
\begin{equation}\label{eq:PikR}
\Pi^k_R=\frac{\mathrm{e}^{k^2R^2/2}}{(2\pi)^{3/2}R^3}\int d^3\,x\,\mathrm{e}^{-r^2/2R^2}\,\frac{\sin(kr)}{kr}\,\pi(\vec{x})\,.
\end{equation}

The observable $\Phi^k_R$ corresponds to the Gaussian smearing of the field over a region of size $R$ that registers spherically symmetric fluctuations at scale $k$, the factor $\sin(kr)/kr$ being the zeroth-order component $j_0(kr)$ appearing in the field expansion of the spherical basis  $Y_{lm}(\theta,\varphi)\, j_l(k r)$. The prefactors are adjusted to ensure the smearings are normalized. The nontrivial components of the symplectic form for this subalgebra read:
\begin{equation}
\Omega_{kk}=-\mathrm{i}\left[\Phi^k_R,\Pi^k_R\right]=\frac{\mathrm{e}^{k^2R^2/2}\sinh(k^2R^2/2)}{4\pi^{3/2}k^2R^5}\,,
\end{equation}
\begin{equation}
\Omega_{kk'}=-\mathrm{i}\left[\Phi^k_R,\Pi^{k'}_R\right]=\frac{\mathrm{e}^{k^2R^2/2}\mathrm{e}^{k'^2R^2/2}(\mathrm{e}^{-\frac{1}{4}(k-k')^2R^2}-\mathrm{e}^{-\frac{1}{4}(k+k')^2R^2})}{8\pi^{3/2}k^2R^5}\,,
\end{equation}
The correlators in the vacuum state read:
\begin{equation}
\langle \Phi^k_R \Phi^k_R\rangle =\frac{2 k^3 R^3 \, _2F_2\left(1,1;2,\frac{5}{2};k^2 R^2\right)+3 \sqrt{\pi } e^{k^2 R^2} \text{erf}(k R)-6 k R}{48 \pi ^2 k^3 R^5}
\end{equation}

\begin{equation}
\langle \Pi^k_R  \Pi^k_R\rangle = 
\frac{e^{k^2 R^2} \text{erf}(k R)}{16 \pi ^{3/2} k R^5}
\end{equation}

\begin{align}
\langle \Phi^k_R \Phi^{k'}_R\rangle &=-\frac{1} {96 \pi ^2 k k' R^5}\Bigg[R^3 
   \left((k-k')^2 \,
   _2F_2\left(1,1;2,\frac{5}{2};\frac{1}{4} (k-k')^2 R^2\right)\right. \nonumber\\
    &\left.-(k+k')^2 \,
   _2F_2\left(1,1;2,\frac{5}{2};\frac{1}{4} (k+k')^2 R^2\right)\right)+\frac{12
   \sqrt{\pi } e^{\frac{1}{4} R^2 (k-k')^2} }{ (k-k') (k+k')}
   \nonumber\\
  &\times \left((k+k')
   \text{erf}\left(\frac{1}{2} R (k-k')\right)-(k-k') e^{kk' R^2}
   \text{erf}\left(\frac{1}{2} R (k+k')\right)\right)\Bigg]
\end{align}

\begin{equation}
\langle \Pi_R^k \Pi_R^{k'}\rangle = \frac{(k+k') e^{\frac{1}{4} R^2 (k+k')^2} \text{erf}\left(\frac{1}{2} R (k+k')\right)-(k-k') e^{\frac{1}{4} R^2 (k-k')^2}
   \text{erf}\left(\frac{1}{2} R (k-k')\right)}{32 \pi ^{3/2} k k' R^5}
\end{equation}

\begin{figure}[t]
\begin{center}
\includegraphics[width=0.7\textwidth]{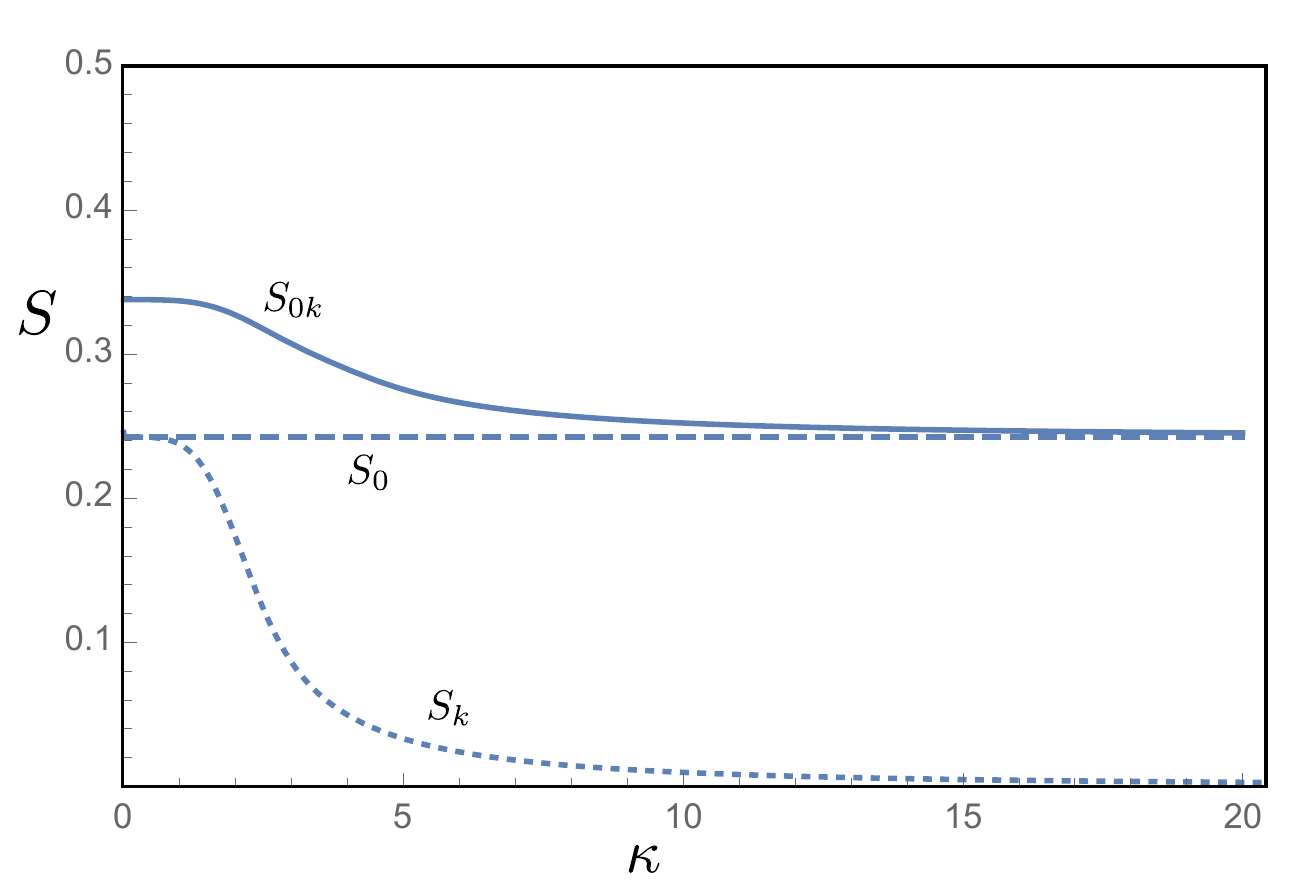}
\end{center}
\caption{Entropy of the $S_{0k}$ subsytem (solid line), the $S_{0}$ subsystem (dashed line), and the $S_{k}$ subsystem (dotted line), as a function of $\kappa=k R$. }
\label{fig:twomode} 
\end{figure}

Using these formulae, we can compute the entropy associated to this subalgebra for any truncation $n$ and any choice of the frequencies $k_j$. As an example, we compute the entropy for the $n=2$ case where $k_1=0,k_2=k$. This case corresponds to enlarging the  subalgebra of the previous subsection, the one generated by (\ref{eq:PhiR})--(\ref{eq:PiR}) (for a massless field), and adding to the generators the observables (\ref{eq:PhikR})--(\ref{eq:PikR}). This entropy $S_{0k}$ can be compared to the already computed entropy $S_0$ of the subalgebra generated by (\ref{eq:PhiR})--(\ref{eq:PiR}) alone, and to the entropy $S_k$ of the subalgebra generated by  (\ref{eq:PhikR})--(\ref{eq:PikR}) alone. The three results are plotted in Figure \ref{fig:twomode} as a function of the dimensionless parameter $\kappa=kR$.

Figure \ref{fig:twomode} shows that the entropy of the combined subsystem is smaller than the sum of the individual entropies.
  In addition, we see that the entropy $S_k$ vanishes for large $\kappa$. In this limit, the high-frequency modes captured by the subalgebra  cannot distinguish between the vacuum state for the field in the whole spacetime (in which different modes are unentangled) and the vacuum restricted to the region of size $R\gg k^{-1}$.

We can enlarge the subalgebra more and more, including as well smearing functions with nontrivial angular dependence, so to capture all the degrees of freedom in a region of size $R$. In such a limit, one can expect that the entropy of the subalgebra approaches the geometric entropy and scales with the area of the region. However, there are two issues involved in taking such a limit:
\begin{itemize}
\item[(i)] The first is the practical necessity of finding a suitable set of observables in which the off-diagonal commutators and correlators (e.g. quantities like $\left[\Phi^k_R,\Pi^{k'}_R\right]$ or $\langle \Phi^k_R \Phi^{k'}_R\rangle$ above) vanish. This is because the computational complexity of diagonalizing a  $2n\times2n$ matrix in the $n\to\infty$ limit becomes prohibitive. 
\item[(ii)] The second issue is that we would like to find observables defined by smearing functions that are strictly zero outside of our $R$-sized region, rather than a smearing function with Gaussian tails outside the region; moreover, the smearing has to be smooth enough so that all correlation functions are well-defined.
\end{itemize}
In the next section we  introduce a basis of observables satisfying these desiderata, and use it to rederive within our framework the area law for the entropy of a spherical region in Minkowski space.

\section{Smearing functions and observables with compact support in a sphere}
\label{sec:sphere}
In this section we introduce a set of smeared observables with compact support in a sphere.  In the appropriate limit, this set is suitable for recovering the area law scaling of the geometric entropy associated with a spherical region in the Minkowski vacuum of a massless scalar field. First of all, we explain how the symmetry properties of the vacuum select a particular complete basis of field fluctuations inside the sphere as the modes that diagonalize the entanglement Hamiltonian and make manifest its thermality. Then, we introduce a discrete set of modes (and the smeared field observables associated to them). These modes approach the thermal modes in a suitable limit. In the next section we then compute the entanglement entropy associated to our discrete set of smeared observables, and show that its scaling recovers the area law as a complete spherical basis is approached.

\subsection{Thermal vacuum and conformal transformations}
\label{sec:thermal}

Besides being Poincar\'e invariant, Minkowski space is also invariant under a conformal transformation which preserves the boundary of a space-like sphere and its causal development \cite{Casini:2010kt}. 

The causal domain of a sphere of radius $R$ is the spacetime region $r+|t|\leq R$. In this region the Minkowski line element can be written as
\begin{align}
ds^2=&\;-dt^2+dr^2+r^2\,\big(d\theta^2+(\sin\theta)^2d\phi^2\big)\\
=&\;\;\Omega(\lambda,\sigma)^2 \;\Big(\!-d\lambda^2+d\sigma^2+(\sinh\sigma)^2\big(d\theta^2+(\sin\theta)^2d\phi^2\big)\Big)\,,
\label{eq:sphereMink}
\end{align}
where  $-\infty<\lambda< \infty$, $0\leq \sigma<\infty$ and the conformal factor $\Omega(\lambda,\sigma)$ is given by
\begin{equation}
\Omega(\lambda,\sigma)=\frac{R}{\cosh\lambda+\cosh\sigma}\,.
\end{equation}
The coordinate transformation from spherical coordinates $(t,r,\theta,\phi)$ to coordinates $(\lambda,\sigma,\theta,\phi)$ is given by 
\begin{equation}
t(\lambda,\sigma)=\Omega(\lambda,\sigma)\;\sinh \lambda\,,\qquad
r(\lambda,\sigma)=\Omega(\lambda,\sigma)\;
\sinh \sigma\,.
\label{eq:rt}
\end{equation}
The expression (\ref{eq:sphereMink}) of the Minkowski metric $\eta_{\mu\nu}$ makes its conformal symmetries manifest, in particular its conformal invariance under shifts of the time-like coordinate $\lambda$. 

It is well known that the restriction of the massless Minkowski vacuum state to the interior of a sphere results in a thermal state
 \cite{Casini:2010kt}. The restriction of the vacuum is thermal due to the $2\pi$-periodicity of the metric (and consequently, the vacuum two-point function) in the imaginary extension of the coordinate $\lambda$. This is analogous to the thermality of the restriction of the vacuum to half space, which is related to the periodicity of the Rindler time coordinate $\eta$ (the time along orbits of the boost Killing vector). In the Rindler case, the basis of modes that expand the field in the half-space that diagonalizes the thermal vacuum is positive frequency in $\eta$. In a similar way, the modes that diagonalize the vacuum restricted to the sphere and make its thermal nature manifest are positive frequency in $\lambda$. We proceed now to find these modes.

\subsection{Orthonormal functions with compact support in a sphere}
\label{sec:orthonormal}
Let us consider a spherical region of radius $R$, and  adopt spherical coordinates $(r,\theta,\phi)$. We consider the transformation
\begin{equation}
\textstyle 
r(\sigma)=R\,\tanh\frac{\sigma}{2}\,,
\end{equation}
and its inverse $\sigma=\sigma(r)$ which maps $r\in (0,R)$ in the semi-infinite domain $\sigma\in (0,\infty)$; this is the same conformal coordinate introduced above in (\ref{eq:rt}), specialized to $\lambda = 0$. We consider next the Laplacian $\Delta_h$ on the constant curvature space with line element $dh^2=d\sigma^2+(\sinh{\sigma})^2\,(d\theta^2+(\sin\theta)^2d\phi^2)$, and define the orthonormal functions $f_{\kappa lm}(r,\theta,\phi)$ as solutions of the differential equation
\begin{equation}
-\Delta_h\, f_{\kappa lm}(r(\sigma),\theta,\phi)=(\kappa^2+1)\;f_{\kappa lm}(r(\sigma),\theta,\phi)\,.
\label{eq:elliptic1}
\end{equation}
These functions have the form
\begin{equation}
f_{\kappa lm}(r,\theta,\phi)=\mathcal{R}_{\kappa l}(r)\,Y_{lm}(\theta,\phi)
\end{equation}
where $Y_{lm}(\theta,\phi)$ are spherical harmonics and the radial functions $\mathcal{R}_{\kappa l}(r)$ have compact support in $r\in (0,R)$. The spacetime modes $\mathrm{e}^{-i\kappa \lambda} f_{\kappa lm}(r,\theta,\phi)$ (suitably normalized) provide a complete orthonormal basis for the field in the sphere's causal domain, and are the modes that diagonalize the entanglement Hamiltonian restricted to the sphere, analogously to the Rindler modes for half-space.   

Note that the index $\kappa$ is continuous, while we are looking for a discrete set so to define a discrete subalgebra  of field observables associated to a range of modes and compute its entropy. We therefore seek a modification of these modes that defines a discrete set such that the continuum limit can be approached in a controlled way.

We define the discrete set of orthonormal functions $f_{nlm}(r,\theta,\phi)$ as solutions of the differential equation
\begin{equation}
\Big(-\Delta_h\;+\;c_0^{\,2}\, \theta(\sigma-\sigma_0)\Big)\, f_{nlm}(r(\sigma),\theta,\phi)=(\kappa^2+1)\;f_{nlm}(r(\sigma),\theta,\phi)\,,
\label{eq:elliptic2}
\end{equation}
where the potential step $c_0^{\,2}\, \theta(\sigma-\sigma_0)$ with $c_0>0$ defines a spherical region of radius $\sigma_0$ and results in a discrete set of eigenvalues $\kappa_{nl}$ for $\kappa\leq c_0$. 
The functions $f_{nlm}(r,\theta,\phi)$ are orthonormal with respect to 
a spherically-symmetric integration measure $q(r)^3\,r^2dr\,\sin\theta\,d\theta d\phi$, i.e.,
\begin{equation}
\int_0^R f_{nlm}(r,\theta,\phi)\,f_{nlm}^*(r,\theta,\phi)\;\,q(r)^3\,r^2dr\,\sin\theta\,d\theta d\phi\;=\delta_{nn'}\delta_{ll'}\delta_{mm'}\,,
\label{eq:ortonormf}
\end{equation}
with the choice $q(r(\sigma))=\frac{2}{R}\big(\!\cosh\frac{\sigma}{2}\big)^2$. This makes the integration measure reduce to
\begin{equation}
q(r(\sigma))^3\,r^2dr\,\sin\theta\,d\theta d\phi\;=\;(\sinh \sigma)^2\,d\sigma\,\sin\theta\,d\theta d\phi\,,
\end{equation}
which is the invariant measure on a constant-curvature space. Note that for large $\sigma_0$ we can define a small distance $\varepsilon$ from the boundary of the sphere,
\begin{equation}
\varepsilon=R-r(\sigma_0)\;\approx\;2R\,e^{-\sigma_0}\,.
\end{equation}
As the small distance $\varepsilon$ is taken close to zero, the potential step in the differential equation defining the modes is removed (going to infinity in the hyperbolic conformal space) and the continuum of exact solutions to the field equation is recovered.

 The $\varepsilon$ parameter plays the role of effective UV cutoff in the computation of the entropy. Its role in the computation is similar to the cutoff in the brick wall regularization of black hole entropy 
  \cite{tHooft:1984kcu}. However, it is conceptually different in two ways. Firstly, due to the finiteness of the potential barrier, it selects modes that vanish smoothly at the boundary of the sphere rather than sharply at a  ``wall'' close to the boundary. Secondly, it will be used to define a discrete set of \textit{observables} (defined by smearing the field with the discrete set of modified mode solutions) without modifying in any way the \textit{theory} or the \textit{quantum state}, as the usual forms of entropy regularization do. This point is expanded upon in Section \ref{sec:observables}.

\subsection{Radial profile of the smearing functions}
In order to determine the radial part $\mathcal{R}_{nl}(r)$ of the smearing function, we consider the change of variables 
\begin{equation}
\mathcal{R}_{nl}(r(\sigma))=\frac{\psi_{nl}(\sigma)}{\sinh{\sigma}}\,,
\end{equation}
which allows us to write the orthonormality condition (\ref{eq:ortonormf}) as
\begin{equation}
\int_0^\infty \psi_{nl}(\sigma)\,\psi_{n'l'}^*(\sigma)\,d\sigma\;=\;\delta_{nn'}\delta_{ll'}\,,
\end{equation}
and the differential equation (\ref{eq:elliptic2}) as
\begin{equation}
-\psi_{nl}''(\sigma)+V_l(\sigma)\,\psi_{nl}(\sigma)=\kappa^2\;\psi_{nl}(\sigma)\,,
\label{eq:Schr}
\end{equation}
with the effective radial potential
\begin{equation}
V_l(\sigma)=\frac{l(l+1)}{(\sinh\sigma)^2}\;+\;c_0^{\,2}\; \theta(\sigma-\sigma_0)\,.
\end{equation}
We have therefore reduced the problem to the one of computing eigenfunctions of a time-independent Schr\"odinger equation. The classical motion in the potential $V_l(\sigma)$ is bounded for $\kappa\leq c_0$. As a result the eigenvalues $\kappa$ are quantized, i.e., they assume only a discrete set of values
\begin{equation}
\kappa=\kappa_{nl} \qquad \mathrm{with} \qquad n=0,1,2,\ldots, N\,.
\end{equation}
We focus on this discrete part of the spectrum. The half-line  $\sigma\in (0,\infty)$ can be divided in three regions: 
\begin{itemize}
\item[(I)] A classically forbidden region $\sigma\in(0,\sigma_\mathrm{min})$ with
\begin{equation}
\sigma_\mathrm{min}=\mathrm{arcsinh}\,\sqrt{\frac{l(l+1)}{\kappa^2}}\,,
\end{equation}
where $\psi(\sigma)$ is exponentially suppressed.
\item[(II)] A classically allowed region $\sigma\in(\sigma_\mathrm{min},\sigma_0)$ where the function oscillates and can be approximated by a WKB wavefunction,
\begin{equation}
\psi_{n l}(\sigma)
\stackrel{{}^{\textrm{WKB}}}{=}
\mathcal{N}_{nl}\;\sin\left( \int_{\sigma_\mathrm{min}}^{\sigma}\!\!\sqrt{\kappa_{nl}^2-\frac{l(l+1)}{(\sinh\sigma')^2}}\;\,d\sigma'\;\,+\;\Theta_{nl}\right)
\end{equation}
where $\mathcal{N}_{nl}$ is a normalization and $\Theta_{nl}$ is fixed by the matching condition with region I.
\item[(III)] A classically forbidden region $\sigma\in(\sigma_0,\infty)$ where the wavefunction decays exponentially.
\end{itemize}
The matching conditions between these three regions result in the Bohr-Sommerfeld quantization condition
\begin{equation}
\left(n+\frac{1}{2}\right)\pi 
\stackrel{{}^{\textrm{WKB}}}{=}
\int_{\sigma_\mathrm{min}}^{\sigma_0}\!\!\sqrt{\kappa^2-\frac{l(l+1)}{(\sinh\sigma)^2}}\;\,d\sigma\;\,,
\label{eq:BohrSomm}
\end{equation}
which is to be understood as an equation for the level $\kappa=\kappa_{nl}$. The discrete level with the largest $n$, denoted $N$ here, is given by $N\stackrel{{}^{\textrm{WKB}}}{=}\lfloor \! n(c_0)\!\rfloor$ where the function $n(\kappa)$ is defined via Eq.~(\ref{eq:BohrSomm}).\footnote{
The integral in Eq.~(\ref{eq:BohrSomm}) can be computed explicitly,
\begin{align*}
n(\kappa)=&\;\frac{\sqrt{l(l+1)}}{\pi}\,\mathrm{arccot}\Bigg(\frac{\kappa\,\mathrm{sech}(\sigma_0)}{\sqrt{l(l+1)}}
\sqrt{(\sinh\sigma_0)^2-\;l(l+1)/\kappa^2\,}\Bigg)-\frac{\sqrt{l(l+1)}}{2}\,+\\
&+
\frac{\kappa}{2\pi}\log\left(\frac{\Big(\cosh\sigma_0+\sqrt{(\sinh\sigma_0)^2-\;l(l+1)/\kappa^2\,}\;\Big)^2}{1\;+\;l(l+1)/\kappa^2}\right)\,.
\label{eq:}
\end{align*}}

\begin{figure}[t]
\begin{center}
\includegraphics[height = 11em]{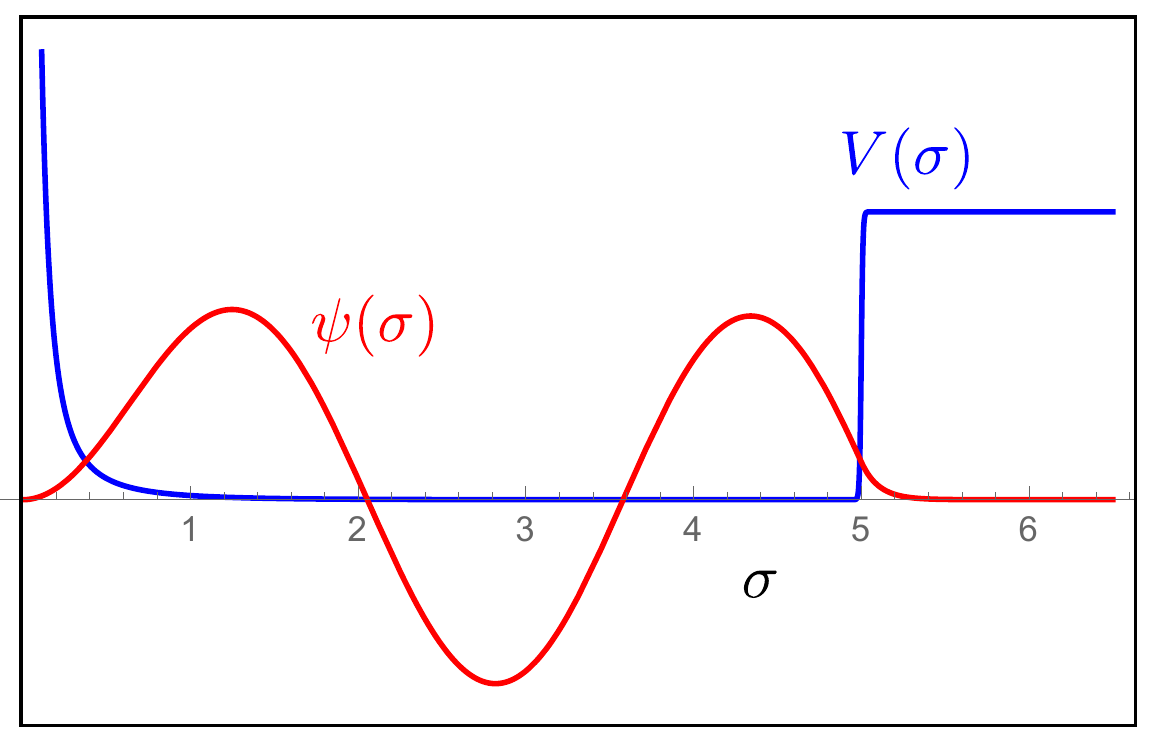}
\hspace{1em}
\includegraphics[height = 11em]{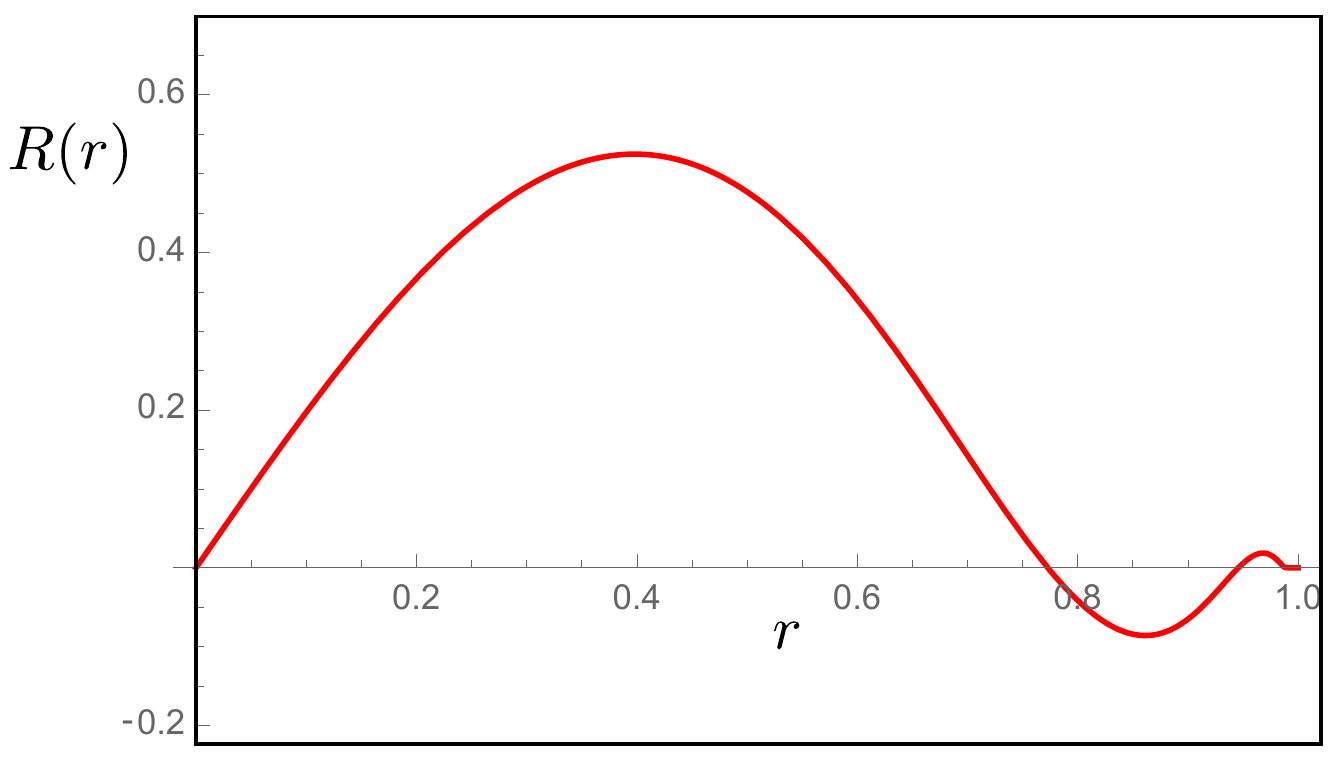}
\caption{Left: Potential $V(\sigma)$ (blue) and eigenfunction $\psi_{21}(\sigma)$ (red). Right: Radial function $\mathcal{R}_{21}(r)$.}
\label{fig:WKB}
\end{center}
\end{figure}

In Figure~(\ref{fig:WKB}) we 
exhibit the potential $V(\sigma)$ and the plot of one particular eigenfunction $\psi_{21}(\sigma)$, together with the associated radial function $\mathcal{R}_{21}(r)=\psi_{21}(\sigma(r))/\sinh(\sigma(r))$. 

The radial functions $\mathcal{R}_{nl}(r)$ show exponential fall-off in the range $r\in(R-\varepsilon,R)$.

\subsection{Density of levels}
\label{sec:density}
In the limit $\sigma_0\to \infty$ with $c_0$ fixed, the number of discrete levels $\kappa_{nl}$ in the interval $[\kappa,\kappa+\delta \kappa]$ diverges and we can define a density of levels $\mu_l(\kappa)$ at fixed $l$. Using the WKB approximation (\ref{eq:BohrSomm}) for the function $n(\kappa)$, we find
\begin{equation}
\mu_l(\kappa)\equiv\frac{dn}{d\kappa}
\stackrel{{}^{\textrm{WKB}}}{=}
\frac{1}{2\pi}\log\left(\frac{\Big(\cosh\sigma_0+\sqrt{(\sinh\sigma_0)^2-\;l(l+1)/\kappa^2\,}\;\Big)^2}{1\;+\;l(l+1)/\kappa^2}\right)\,.
\end{equation}

\begin{figure}[t]
\begin{center}
\includegraphics[height = 18em]{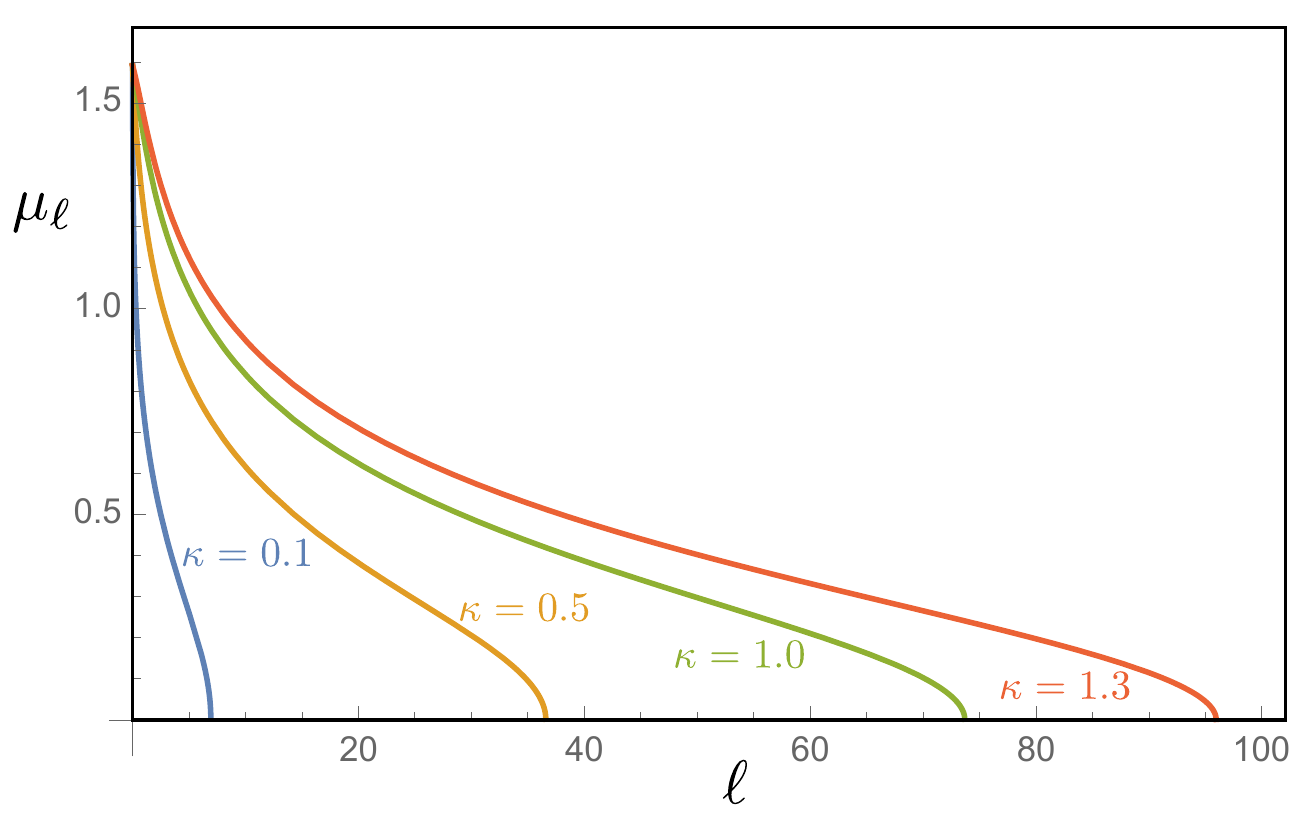}
\caption{Level density at fixed $\kappa$ and $\sigma_0$, as a function of $l$, for four different values of $\kappa$.}
\label{fig:leveldensity}
\end{center}
\end{figure}

The density of levels is plotted as a function of $l$ for fixed $\sigma_0$ and $\kappa$ in Figure~\ref{fig:leveldensity}. 
Note that the density of levels is defined only for $l\leq l_\mathrm{max}$, where
\begin{equation}
l_\mathrm{max}(l_\mathrm{max}+1)=\kappa^2\,(\sinh\sigma_0)^2\,,
\end{equation}
and vanishes at $l_\mathrm{max}$
: $\mu_{l_\mathrm{max}}(\kappa)=0$.

The density of levels allows us to compute sums over $n$ as integrals over $\kappa$ in the limit $\sigma_0\to \infty$,
\begin{align}\label{eq:sumtoint}
\sum_{n}\sum_{l}\sum_{m=-l}^l\qquad &\longrightarrow \qquad\sum_{l}\sum_{m=-l}^l\int d\kappa\;\mu_l(\kappa)\;,\\[1em]
\delta_{nn'}\delta_{ll'}\delta_{m m'}\qquad &\longrightarrow \qquad \frac{1}{\mu_l(\kappa)}\delta(\kappa-\kappa')\,\delta_{ll'}\delta_{m m'}\;.
\end{align}
It is also useful to note that the discrete basis of eigenfunctions $\psi_{nl}(\sigma)$ goes to the continuous basis of radial solutions of (\ref{eq:Schr}) in an infinite domain, given by the Dolginov-Toptygin functions\footnote{
The Dolginov-Toptygin functions (see for example \cite{Lyth:1995cw}) are given by the expression
\begin{equation}
\mathcal{D}^{(-)}_{\kappa l}(\sigma)=\sqrt{\frac{2/\pi}{\prod_{m=0}^l(\kappa^2+m^2)}}\;\;(\sinh\sigma)^l \left(\frac{-1}{\sinh \sigma}\frac{d}{d\sigma}\right)^{\!l+1}\cos(\kappa \sigma)\,.
\end{equation}}
\begin{equation}
\hspace{-.8em}\frac{\psi_{nl}(\sigma)}{\sinh\sigma}\quad\longrightarrow\quad \frac{1}{\sqrt{\mu_l(\kappa)}}\, \;\mathcal{D}^{(-)}_{\kappa l}(\sigma)\,,
\end{equation}
which satisfy Eq.~(\ref{eq:elliptic2}) with $\sigma_0\to \infty$ and are Delta-function orthonormal
\begin{equation}
\int_0^\infty \mathcal{D}^{(-)}_{\kappa l}(\sigma)\;\mathcal{D}^{(-)}_{\kappa l}(\sigma)\;(\sinh\sigma)^2 d\sigma\;=\;\delta(\kappa-\kappa')\,\delta_{ll'}\,.
\end{equation}
Up to a normalization factor, the Dolginov-Toptygin functions are precisely the continuum radial solutions $\mathcal{R}_{\kappa l}(r(\sigma))$ described in Section \ref{sec:orthonormal}.

We note that the limit in which the density of levels diverges,  $\sigma_0\to \infty$, corresponds to a vanishing size of the region $(R-\varepsilon,R)$. 
 The total number of levels in the infinitesimal interval $[\kappa,\kappa+\delta \kappa]$ is given by
\begin{equation}
\sum_{l=0}^{l_\mathrm{max}}(2l+1)\,\mu_l(\kappa)\,\delta \kappa\;
\stackrel{{}^{\textrm{WKB}}}{=}
\;\frac{\sinh(2\sigma_0)-2\sigma_0}{2\pi}\;\kappa^2\,\delta \kappa\;\approx\; \frac{2R^2}{\pi\,\varepsilon^2}\;\kappa^2\,\delta \kappa\,,
\label{eq:intdensity}
\end{equation}
which diverges as $(R/\varepsilon)^2$ in the limit $\varepsilon\to 0$.\\

\subsection{Smeared observables in a spherical region}
Having set up the necessary preliminary tools, we now define a set of smeared observables with support in a sphere of radius $R$ as
\begin{align}
\Phi_{nlm}=&\int_0^R \phi(r,\theta,\phi)\;\mathcal{R}_{nl}(r)\,Y_{lm}(\theta,\phi)\;\,q(r)^2\;\,r^2dr\,\sin\theta\,d\theta d\phi\,,\\[1em]
\Pi_{nlm}=&\int_0^R \pi(r,\theta,\phi)\;\mathcal{R}_{nl}(r)\,Y_{lm}(\theta,\phi)\;\,q(r)\;\,r^2dr\,\sin\theta\,d\theta d\phi\,.
\label{eq:}
\end{align}
The smearing functions for the field 
$\phi(r,\theta,\phi)$ and the momentum $\pi(r,\theta,\phi)$ vanish at the boundary of the sphere and fall off to zero exponentially in the region $r\in(R-\varepsilon,R)$. 
 The observables  satisfy canonical commutation relations
\begin{equation}
[\Phi_{nlm},\Pi_{n'l'm'}]=\mathrm{i}\,\delta_{nn'}\delta_{ll'}\delta_{m,-m'}\,,
\label{eq:diagcomm}
\end{equation}
which follow from the orthonormality of the smearing functions with respect to the integration measure $q(r)^3$, specifically
\begin{equation}
\int_0^R \mathcal{R}_{nl}(r)\;\mathcal{R}_{n'l}(r)\;\,q(r)^3\;\,r^2dr\;=\;\int_0^\infty \psi_{nl}(\sigma)\;\psi_{n'l}(\sigma)\;d\sigma\;=\;\delta_{nn'}\,.
\end{equation}
The observables $\Phi_{nlm}$ and $\Pi_{nlm}$ are defined for $n=1,\ldots,N$, with $N\to \infty$ as $\varepsilon\to 0$.

We have shown that these observables satisfy the second desideratum (ii) listed at the end of Section \ref{sec:vacuum}: they strictly vanishing outside the spherical region and they are smooth enough to guaranty that the correlation functions are finite. These observables satisfy, in part, also the first desideratum (i): the off-diagonal commutators vanish by construction and off-diagonal correlation functions, while not identically zero when evaluated in the Minkowski vacuum, they vanish in the limit in which the continuum basis is recovered. As explained in Section \ref{sec:thermal}, this is a manifestation of the diagonal and thermal nature of the entanglement Hamiltonian in the continuum basis approached by the discrete modes in the limit $\varepsilon\to 0$. In this limit, the correlation functions of the observables $\Phi_{nlm}$ and $\Pi_{nlm}$ take the simple form:
\begin{align}\label{spheroncorr1}
& \langle 0|\Phi_{nlm}\,\Phi_{n'l'm'}|0\rangle\;\approx\;\frac{1}{4\pi \,\kappa_{nl}\,\tanh (\pi\kappa_{nl})}\,\delta_{nn'}\delta_{ll'}\delta_{m,-m'}\,,\\[1em]
& \langle 0|\Pi_{nlm}\,\Pi_{n'l'm'}|0\rangle\;\approx\;\frac{\kappa_{nl}}{4\pi\,\tanh (\pi\kappa_{nl})}\,\delta_{nn'}\delta_{ll'}\delta_{m,-m'}\,,\\[1em]
&  \langle 0|\Phi_{nlm}\,\Pi_{n'l'm'}+\Pi_{n'l'm'}\,\Phi_{nlm}|0\rangle=0\,.
\label{spheroncorr3}
\end{align}
Mode by mode, these are the correlation functions of a thermal harmonic oscillator of frequency $\kappa_{nl}$ and temperature $1/2\pi$. The $\approx$ notation implies equality up to corrections that vanish in the limit $\varepsilon\to 0$

\section{Entanglement entropy of observables in a spherical region}
\label{sec:observables}
With all the pieces in place, the computation of the entanglement entropy of a subalgebra of smeared field observables is a simple matter. The diagonal  commutators (\ref{eq:diagcomm}) and the diagonal correlators (\ref{spheroncorr1})--(\ref{spheroncorr3}) imply that the restricted complex structure $\mathrm{i}J_A$ for the subalgebra of observables has eigenvalues 
\begin{equation}
\nu_{nl}=\left(\tanh  (\pi\kappa_{nl}) \right)^{-2}\,.
\label{eq:spheronnu}
\end{equation}
The entanglement entropy of modes in the range $\kappa_{nl}\in [\kappa_{\textrm{min}},\kappa_{\textrm{max}}]$ is
\begin{equation}
S_A(|0\rangle)=\sum_{nlm}s(\kappa_{nl})\;=\; \sum_{l=0}^{\infty}(2l+1)\, \sum_{n=n_{\textrm{min}}\!\!\!}^{n_{\textrm{max}}}\, s(\kappa_{nl})\\[1em]
\end{equation}
where 
\begin{equation}
s(\kappa)=-\ln\left(1-\mathrm{e}^{-2\pi\kappa}\right)+2\pi\kappa\frac{\mathrm{e}^{-2\pi\kappa}}{1-\mathrm{e}^{-2\pi\kappa}}
\label{eq:}
\end{equation}
is the result of (\ref{eq:snu}) applied to (\ref{eq:spheronnu}). This result coincides with the entropy of an oscillator at temperature $T=2\pi$.

In the $\varepsilon\to 0$ limit we can evaluate the entropy using our results on the density of levels obtained in Section \ref{sec:density}. Using (\ref{eq:sumtoint}) and (\ref{eq:intdensity}), we get:
\begin{align}
S_A(|0\rangle)&\approx \sum_{l=0}^{\infty}(2l+1)\,\int_0^\infty s(\kappa)\;\mu_l(\kappa)\;d\kappa\;\approx \;\int_{\kappa_{\textrm{min}}}^{\kappa_{\textrm{max}}}s(\kappa)\;  \left(\frac{2R^2}{\pi\,\varepsilon^2}\;\kappa^2\right) d\kappa\\[1em]
& = c(\kappa_{\textrm{min}},\kappa_{\textrm{max}})\;\,\frac{\textrm{Area}(R)}{\varepsilon^2}\,,
\end{align}
where $\textrm{Area}(R)=4\pi \,R^2\;$ and
\begin{equation}
c(\kappa_{\textrm{min}},\kappa_{\textrm{max}})=\frac{1}{4\pi}\int_{\kappa_{\textrm{min}}}^{\kappa_{\textrm{max}}}s(\kappa)\;  \left(\frac{2R^2}{\pi\,\varepsilon^2}\;\kappa^2\right) d\kappa\,.
\end{equation}
The entropy of a subalgebra of observables capturing any finite range of radial modes (and all angular degrees of freedom in that range) is therefore proportional to the area of the sphere divided by the cutoff parameter $\varepsilon^2$. (Recall that the radial frequencies $\kappa$ are dimensionless and the dimensionful parameter $\varepsilon$ guaranties that the smearing functions vanish smoothly at the boundary of the sphere). In the limit $\kappa_{\textrm{min}}\to 0,\,\kappa_{\textrm{max}}\to \infty$, this entropy approaches the entanglement entropy of all the degrees of freedom in the spherical region of radius $R$. To the leading order in the parameter $\varepsilon$, this is given by:
\begin{equation}\label{finalresult}
S_R(|0\rangle)=\lim_{\substack{\kappa_{\textrm{min}}\to 0\\ \kappa_{\textrm{max}}\to \infty}} S_A(|0\rangle)\;=\;\frac{1}{360\pi}\frac{\textrm{Area}(R)}{\varepsilon^2}+\ldots
\end{equation}
We have therefore recovered the area law for the geometric entropy.

\medskip

In calculations of the geometric entanglement entropy it is known that the numerical coefficient in front of the area law is not universal as it depends on the specific regularization method employed. Intriguingly, in our result (\ref{finalresult}) we find a coefficient $1/360\pi$ which matches the one appearing in the brick-wall regularization \cite{tHooft:1984kcu}  of the entanglement entropy across a planar surface of Minkowski half-space found in  \cite{Susskind:1994sm}. We note that, while these coefficients turn out to be the same, the intermediate steps of the calculation differ. More importantly, the brick-wall regularization modifies the vacuum state in the vicinity of the boundary of the sphere, while our construction never modifies the state: it is the subalgebra of observables that probes the state only with finite resolution (measured by the parameter $\varepsilon$) therefore rendering the entropy finite.

\section{Discussion}
\label{sec:discussion}
Measurements of a field are often restricted to a region of space and have only a finite resolution. Such measurements can be described as a subalgebra of observables generated by a linear smearing of the field against smooth functions with support on the region. The uncertainty in the results of such measurements is characterized by the entropy of the state restricted to the subalgebra. In Section \ref{sec:Gaussian}, we showed how to compute the entropy of a Gaussian state restricted to such a subalgebra using linear symplectic methods adapted from \cite{holevo2013quantum,weedbrook2012gaussian,adesso2014continuous,Bianchi:2017kgb,Bianchi2017kahler}. Using this definition, in Section \ref{sec:vacuum} we computed the entropy of the Minkowski vacuum state restricted to some simple classes of smeared field observables.

The geometric entanglement entropy can be understood as the entropy of the vacuum state of a field theory, restricted to a region of space. The result of this calculation is generally divergent because of the presence of short-ranged correlations across the boundary of the region of space. A standard procedure for defining the geometric entropy involves a modification of the field theory in the short-ranged correlations of the field theory via the introduction of a UV cutoff. The result of this procedure is an area law for the geometric entropy. Here we proposed a different, operational definition of the geometric entropy that does not involve a modification of the theory in the UV. A set of measurements with finite resolution  provides a subalgebra of observables that does not probe short-ranged correlations, and therefore there is no need to introduce ad hoc modifications of the theory or the state in the UV. The choice of subalgebra is dictated by the set of observables we measure. In particular, in Section \ref{sec:sphere} and \ref{sec:observables} we considered the entropy of the Minkowski vacuum of a massless scalar field restricted to a spherical region. In order to provide a concrete example we considered  a specific subalgebra of observables and showed that refining it and increasing its resolution, the standard formula for the area law is recovered.

Identifying a subalgebra of observables that can be easily refined, while keeping the computation feasible, is non-trivial. Here we started by considered smearing functions that formally diagonalize the modular Hamiltonian in a spherical region. Such functions are eigenfunction of a specific differential operator and form a continuous set labeled by a radial quantum number $\kappa$. In order to extract a finite set of smearing functions that vanish smoothly at the boundary of the sphere, we introduced a step function at distance $\varepsilon$ from the boundary, which results in a quantization $\kappa
_{n\ell}$ of $\kappa$. In the limit  $\varepsilon\to 0$, the full subalgebra associated with the interior of the sphere is recovered, and the entropy of the subalgebra is found to approach the divergent geometric entropy with an area law.

\section*{Acknowledgments}
We thank Ivan Agullo, Abhay Ashtekar, Lucas Hackl, Ted Jacobson and Rafael Sorkin for useful discussion. The work of E.B. is supported by the NSF Grant PHY-1806428.

\end{document}